\documentclass[10pt, twocolumn, english, amssymb, aps, prx, superscriptaddress, showpacs, showkeys, floatfix, longbibliography]{revtex4-2}

\usepackage{booktabs}
\usepackage{amsmath, amssymb, graphicx}
\usepackage{mathptmx}
\usepackage{physics}
\usepackage{xcolor}
\usepackage{textcomp}
\usepackage[utf8]{inputenc}
\usepackage[english]{babel}
\usepackage[scr=rsfso, cal=zapfc, frak=euler, bb=ams]{mathalfa}
\usepackage{enumitem}
\usepackage{tikz}
\usetikzlibrary{arrows.meta, positioning, calc}
\usepackage{array,ragged2e}

\makeatletter
\@ifundefined{l@en}{\let\l@en\l@english}{}
\makeatother

\usepackage{booktabs}
\usepackage{amsmath, amssymb, graphicx}
\usepackage{mathptmx}
\usepackage{physics}
\usepackage{xcolor}
\usepackage{textcomp}
\usepackage[utf8]{inputenc}
\usepackage[scr=rsfso, cal=zapfc, frak=euler, bb=ams]{mathalfa}
\usepackage{enumitem}
\usepackage{tikz}
\usetikzlibrary{arrows.meta, positioning, calc}
\usepackage{array,ragged2e}

\usepackage{hyperref}

\newcolumntype{Y}{>{\RaggedRight\arraybackslash}X}

\usepackage{subfiles}

\begin{document}

\title{
	Parametrically driven Kerr temporal soliton crystals 
}

\author{Yifan Sun}
\email{Corresponding author: yifan.sun@ulb.be}
\affiliation{Service OPERA-Photonique, Université Libre de Bruxelles, 50 Av. F. D. Roosevelt, B-1050 Brussels, Belgium}	
\author{Clément Dupont}
\affiliation{Service OPERA-Photonique, Université Libre de Bruxelles, 50 Av. F. D. Roosevelt, B-1050 Brussels, Belgium}		
\affiliation{Laboratoire d'Information Quantique, Université Libre de Bruxelles, B-1050 Bruxelles, Belgium}
\author{Edem Kossi Akakpo}
\affiliation{Service OPERA-Photonique, Université Libre de Bruxelles, 50 Av. F. D. Roosevelt, B-1050 Brussels, Belgium}			
\author{Francesco De Lucia}
\affiliation{Service OPERA-Photonique, Université Libre de Bruxelles, 50 Av. F. D. Roosevelt, B-1050 Brussels, Belgium}		
\affiliation{CNRS, UMR 8523 – PhLAM, Université de Lille, F-59000 Lille, France}
\author{Georges Semaan}
\affiliation{Service OPERA-Photonique, Université Libre de Bruxelles, 50 Av. F. D. Roosevelt, B-1050 Brussels, Belgium}		
\author{Simon-Pierre Gorza}
\affiliation{Service OPERA-Photonique, Université Libre de Bruxelles, 50 Av. F. D. Roosevelt, B-1050 Brussels, Belgium}		
\author{François Leo}
\affiliation{Service OPERA-Photonique, Université Libre de Bruxelles, 50 Av. F. D. Roosevelt, B-1050 Brussels, Belgium}

\begin{abstract} 
	
    We theoretically investigate the dynamics of parametrically driven soliton crystals (PDSC) and their associated frequency combs in doubly resonant cavities with quadratic and cubic nonlinearities. 
    We show that, in a regime with strong pump--signal walk-off where the homogeneous state is unstable, noise-seeded dynamics relaxes toward stable, equally spaced multi-soliton states. At fixed detuning, the driving strength acts as the primary control parameter for the soliton number.
    Pump–signal walk-off extends pump depletion from a local perturbation to a global constraint, enabling long-range soliton interactions; in combination with the pump phase, this mechanism stabilizes the crystal’s equal spacing and preserves comb coherence.
	Additionally, we identify a novel nonlinear state in parametric soliton crystals in which circulating solitons periodically alternate their intensities and group velocities, a phenomenon we term the {\it soliton-pursuing} state.
    Furthermore, due to the phase-selective nature of the optical parametric process, we show how different configurations of soliton phases determine the optical frequency combs, enabling odd-harmonic and subharmonic-like combs.
\end{abstract}

\maketitle
{\it Introduction.} 
{
Self-organized patterns \cite{crossPatternFormationOutside1993} in coupled dynamical systems underpin emergent behaviors across diverse disciplines, from social networks \cite{Mariani2024} and biological cells \cite{Kawaguchi2017} to bacterial colonies \cite{Sokolov2007} and topological materials \cite{Li2021}.
These systems exhibit emergent patterns often driven by nonlinear interactions among individual components \cite{Casadiego2017}. Self-organization also occurs in optical systems, where nonlinear wave interactions give rise to a wide range of complex states \cite{Dudley2014}.
Cavity solitons (CS) \cite{Wabnitz1993,Leo2010,Herr2014}, a special class of dissipative solitons \cite{grelu_dissipative_2012}, arise in optical resonators from the balance between dispersion and nonlinearity, with coherent driving compensating dissipation \cite{haelterman_dissipative_1992,Wabnitz1993}.
They occur in both spatial and temporal settings \cite{kartashov_frontiers_2019,sun2024multidimensional,Sun2023bullets,sunTransdimensionalDynamicsKerr2025}, making optical cavities compelling platforms for exploring nonlinear many-body interactions \cite{Rotschild2006,Peleg2014}.
CS form optical frequency combs (OFC), which have revolutionized modern photonics, driving advancements in high-speed communications \cite{MarinPalomo2017}, low-noise radiofrequency generation \cite{Lucas2020}, and precision distance ranging \cite{Trocha2018}. Investigations into soliton stability and dynamics have revealed a variety of solitonic states, including breather solitons \cite{Guo2017,Yu2017,Lucas2017}, bound soliton molecules \cite{Liu2023,Weng2020}, and localized chaotic soliton interactions \cite{Nielsen2021,Sun2022OL,Sun2023chaos}.
Beyond individual solitons, nonlinear optical cavities can support soliton crystals (SC), which are periodic arrays of solitons within a cavity. SC provide new opportunities for enhancing the power efficiency and tunability of OFC while preserving coherence \cite{Cole2017,Karpov2019,He2020,Lu2021,taheri2022,Guidry2023,Hu2024}.

To date, CS have been extensively studied in Kerr-type cavities, where cubic ($\chi_3$) nonlinearity dominates their formation. However, recent advancements have focused on quadratic ($\chi_2$) nonlinear systems, which expand solitonic OFC generation through processes such as second-harmonic generation (SHG) \cite{leo_frequency-comb_2016,Leo2016,Hansson2017,Xue2017,Xue2018,Villois2019a,MasArabi2020,Arabi2020,ParraRivas2021,mas_arabi_depletion-limited_2023,talentiInterplayH2H32025,talentiBistableSolitonOptical2025} and optical parametric oscillation (OPO) \cite{Longhi1996,Amiune2021,Villois2019,Jankowski2018,Roy2021,Englebert2021a}. 
In degenerate OPO, $\chi_2$ nonlinearity enables the conversion of pump photons at $2\omega$ into signal and idler photons at $\omega$, allowing CS at $\omega$ to be sustained by parametric pumping at $2\omega$. These solitons, known as parametrically driven cavity solitons (PDCS) \cite{Englebert2021a}, can exist in either an in-phase or out-of-phase state (0 or $\pi$). {\color{black}This binary phase behavior in degenerate OPO has been used for studying Ising machines~\cite{takedaBoltzmannSamplingXY2017,wangCoherentIsingMachine2013}, where each phase represents a spin state, and initial quantum noise seeds the random phase selection~\cite{marandiAllopticalQuantumRandom2012}.} 

PDCS have been experimentally observed in both fiber ring cavities \cite{Englebert2021a} and microcavities \cite{Bruch2021}. 
Depending on whether the pump field is resonant within the cavity, these systems are classified as singly resonant (only the signal resonates) \cite{mas_arabi_depletion-limited_2023} or doubly resonant \cite{Ding2024}. 
Related integrated $\chi^{(2)}$/$\chi^{(3)}$ platforms, including AlN, SiC, and thin-film LiNbO$_3$~\cite{Bruch2021,wangSolitonFormationSpectral2022,zhengOctavespanningDeterministicSingle2025,heSelfstartingBichromaticLiNbO2019,songOctavespanningKerrSoliton2024}, provide promising settings for exploring parametrically driven nonlinear states.
More broadly, PDCS states have also been demonstrated in pure-Kerr systems through dual-wave pumping \cite{moille_parametrically_2024}. 
Despite recent advances, the formation and dynamics of PDCS, particularly the emergence of complex multi-soliton states, remain largely unexplored.

In this work, we show that parametrically driven multi-soliton states in doubly resonant nonlinear cavities spontaneously self-organize into equidistant soliton crystal states [see Fig.~\ref{fig1_concept_excitation}(a)]. Unlike conventional soliton or soliton-crystal states, which are difficult to access and do not readily self-recover after a perturbation~\cite{rowleySelfemergenceRobustSolitons2022}, PDSCs act as robust attractors in a monostable regime with significant pump–signal walk-off. Bifurcation analysis and simulations reveal that the soliton number is set primarily by the driving strength, with stability enforced by pump depletion and pump phase. Pump–signal walk-off further transforms pump depletion from a local to a global effect, mediating long-range interactions that lock both spacing and phase. We also uncover a previously unreported {\it soliton-pursuing state}, distinct from conventional breathing dynamics, where solitons periodically alternate their intensities and group velocities. {\color{black}Moreover, the binary phase of PDCS endows the soliton crystal with a two-level phase degree of freedom that breaks discrete translation symmetry and generates rich frequency-comb structures, including odd-harmonic combs and subharmonic-like spectra analogous to discrete time crystals.}

{\it Model.} To explore the system dynamics, the model \cite{Bruch2021,Ding2024} governing the dynamics of PDSC can be normalized as [see normalization and physical parameters in Appendix]:
\begin{equation}
    \begin{aligned} 
		\frac{\partial A}{\partial T} 
		&= \left[ 
		-1
		-i\delta
		+i\frac{\partial^2}{\partial t^2}
		\right]A
		+
		i\left[
		(\left|A\right|^2+2|B|^2)A
		+gBA^*
		\right],\\
		\frac{\partial B}{\partial T} 
		&= \left[ -r_\mathrm{l}
		-i\left(2\delta+\delta_0\right)
		-d\frac{\partial}{\partial t}
		+ir_\mathrm{g}\frac{\partial^2}{\partial t^2}
		\right]B\\
		&+i\left[
		r_\mathrm{n}(\left|B\right|^2
		+2\left|A\right|^2)B
		+gA^2
		\right]
		+P,\\
	\end{aligned}
	\label{eq_main}
\end{equation}
where $A$ and $B$ represent the amplitudes of the signal and pump waves, which are resonant at $\omega$ and $2\omega$, respectively.
The parameters are defined as follows: $P$ is the coherent driving at $2\omega$; $\delta$ is the frequency detuning; $\delta_0$ and $d$ represent phase and group velocity mismatches, respectively; $g$ is the quadratic nonlinearity coefficient; and $r_\mathrm{l}$, $r_\mathrm{g}$, and $r_\mathrm{n}$ are the ratios of loss, group velocity dispersion, and nonlinearity between the pump and signal. 
Taking the physical values of AlN microcavities in Ref.~\cite{Ding2024} [see details in Appendix], the normalized cavity values are: $d = 669$, $g = 12.2$, $\delta_0 = 0$, $r_\mathrm{l} = 4$, $r_\mathrm{g} = -1$, and $r_\mathrm{n} = 2$. We can retrieve three important scaling parameters: $t_\mathrm{s} = \rm 41.8\,fs$ (time), $|A_\mathrm{s}| = \rm 4.365\,W^{1/2}$ (amplitude), and $T_\mathrm{s}/T_{\rm R} = 175$ (round trips). In other words, one normalized unit in fast time $t$, amplitudes $A$ (and $B$), and slow time $T$ corresponds to the respective physical values as defined above. The roundtrip time is $T_R/t_\mathrm{s}\approx 66.36$.
To better capture the generality of PDSC dynamics, all parameters and values in this paper are expressed in their normalized form.

\begin{figure}[!b]
	\centering
	\includegraphics[scale=1]{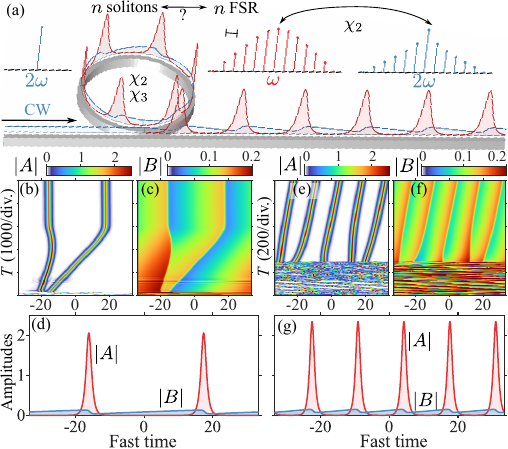}
	\caption{
		(a) Schematic of a soliton crystal in a cavity with $\chi_2$ and $\chi_3$ nonlinearities. 
(b-g) Evolution of $|A(t,T)|$ and $|B(t,T)|$ over time, along with the final profile $|A|$, for driving strengths $P = 2.5$ in (b-d) and $P = 6.5$ in (e-g), with fixed detuning $\delta=2$ throughout [see Movie \href{https://youtu.be/WX0QvjTKGD4}{I} and \href{https://youtu.be/BcQJ2uisnf8}{II}]. (d,g) Final temporal profiles at $T = 5000$ and $1000$.
	}
	\label{fig1_concept_excitation}%
\end{figure}	

{\it Excitation.}  
{
Although detuning scans can facilitate or accelerate access to soliton and PDSC states, stable PDSCs can also form directly at fixed detuning in this cavity.
Figure~\ref{fig1_concept_excitation} shows two examples of soliton crystal generation at fixed detuning $\delta = 2$: 
$P = 2.5$ [Figs.~\ref{fig1_concept_excitation}(b–d)] and $P = 6.5$ [Figs.~\ref{fig1_concept_excitation}(e–g)]\footnote{In calculations, we introduced an additional group delay $d_\mathrm{c}$ in the reference frame for both fields [see Appendix]. This adjustment keeps the soliton field nearly centered in the frame, allowing better visualization and tracking of group velocity variations, without affecting the underlying results. The reference frame delays are $d_\mathrm{c}= 0.3892$ in Fig.~\ref{fig1_concept_excitation}(b--d) and $d_\mathrm{c}= 0.352$ in Fig.~\ref{fig1_concept_excitation}(e--g). 
The two- and five-soliton buildup processes are shown in Movies~\href{https://youtu.be/WX0QvjTKGD4}{I} and \href{https://youtu.be/BcQJ2uisnf8}{II}, respectively.}.
Starting from white noise ($\max|A| = 0.01$), $|A|$ and $|B|$ evolve chaotically for $0 < T < 300$ before forming two solitons [Figs.~\ref{fig1_concept_excitation}(b–c)]. 
Initially close, the soliton distance progressively evolves to equally spaced solitons at $T = 3000$, forming a two-soliton crystal.
The final steady-state profiles [Fig.~\ref{fig1_concept_excitation}(d)] show a sawtooth pump and $\mathrm{sech}$-shaped solitons aligned with the pump edges.  
This shape results from pump depletion, reinforced by pump–signal walk-off, which mediates long-range interactions between solitons; without significant walk-off, pump depletion would remain localized.
For $P = 6.5$, five PDSCs emerge as early as $T = 300$ [Figs.~\ref{fig1_concept_excitation}(e–f)], appearing irregularly before settling into evenly spaced, uniform-intensity patterns. PDSCs act as robust attractors: for fixed driving and detuning, the system evolves into a PDSC, with formation times varying across cases. 
Detuning-scan results with the same parameters as Fig.~\ref{fig1_concept_excitation}, presented in Appendix, show qualitatively similar behavior.

\begin{figure}[!b]
	\centering
	\includegraphics[scale=1]{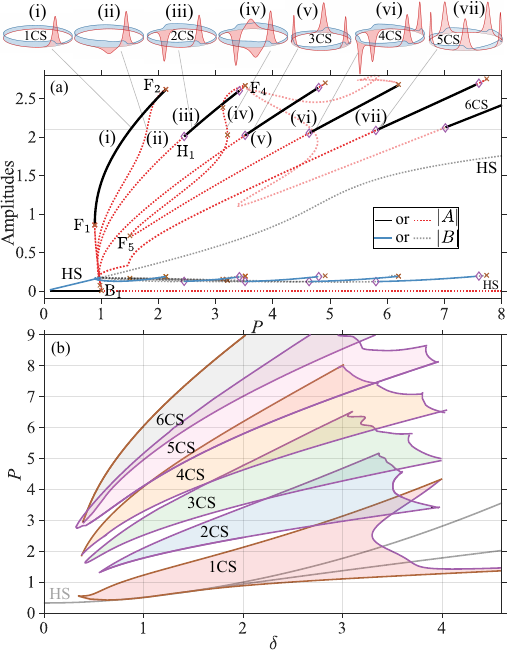}	
	\caption{
		(a) Bifurcation diagram showing 
        $\mathrm{max}(|A|)$ (black and red curves) and 
        $\mathrm{max}(|B|)$ (blue curves), as the driving $P$ is varied, when $\delta=2$. Stable (unstable) states are represented by solid (dashed) curves. The branches (i–vii) correspond to PDSC with different soliton numbers. To illustrate soliton phase relationships in diagrams, only the real part of the amplitude is plotted for each branch [See 
        Movie \href{https://youtu.be/p08_3cc0Og4}{III}].
		(b) Phase diagram of stable PDSC as a function of driving $P$ and detuning $\delta$.
	}
	\label{fig2_diagram}
\end{figure}

{\it Bifurcation and phase diagrams.}  
To better understand the system’s dynamics, we performed path continuation using pde2path\cite{Uecker2014a}, by varying $P$ while fixing $\delta = 2$. 
{\color{black} The steady states for fields $A$ and $B$ satisfy $\partial_T A=\partial_T B=0$; their linear stability is given by the eigenvalues of the Jacobian for small perturbations and is confirmed by the split-step Fourier method.}
The results, shown in Fig.~\ref{fig2_diagram}(a), depict the maximum field amplitudes $\mathrm{max}(|A|)$ 
and $\mathrm{max}(|B|)$ 
as functions of $P$. For small $P$, only the homogeneous solution (HS) is stable, with $|B|$ increasing linearly with $P$ while $|A| = 0$. Beyond the bifurcation point $\rm B_1$ at $P \approx 1$, the HS becomes unstable, giving rise to new branches of localized soliton states.

On branch (i), a single CS state bifurcates from the HS at $\rm B_1$, with its amplitude increasing significantly. The CS stabilizes after crossing fold bifurcation $\rm F_1$ but loses stability upon reaching $\rm F_2$. 
To compare state transitions, Figs.~\ref{fig2_diagram}(i,ii) show the real parts of the amplitudes $\mathrm{Re}(A)$ (red) and $\mathrm{Re}(B)$ (blue), for the solutions marked on branches (i) and (ii). After $\rm F_2$, a side peak with opposite phase appears at the soliton's leading edge [see Fig.~\ref{fig2_diagram}(ii)].  
As the states on branch (ii) approach $\rm B_1$, the anti-phase peak gradually moves farther apart with decreasing amplitudes. Near $\rm B_1$, a fold bifurcation forms equally spaced, out-of-phase solitons. Along branch (iii), their amplitudes grow with increasing $P$ but remain unstable until reaching the Hopf bifurcation $\rm H_1$. 
As $P$ increases, the two solitons undergo similar bifurcations, forming branch (iv), where anti-phase peaks appear ahead of their leading edges [see Fig.~\ref{fig2_diagram}(iv)]. At fold $\rm F_5$, the branch transitions to a four-soliton state [see Fig.~\ref{fig2_diagram}(vi)], with two adjacent solitons in-phase and the other two in anti-phase.  
According to these branches, increasing $P$ doubles soliton number, but can PDSC with an odd number of solitons exist? Applying detuning scans, we obtain PDSC with three ($P = 4$) and five ($P = 6.5$) solitons. Using path continuation, we obtain the corresponding branches in Fig.~\ref{fig2_diagram}(v,vii). 
Numerical simulations confirm these branches by initializing a six-soliton crystal, with each dropping out as the pump gradually decreases [see Fig.~S3 in Appendix and 
Movie \href{https://youtu.be/-vWOQzNRxHg}{IV}]. 

\begin{figure}[!b]
	\centering
	\includegraphics[scale=1]{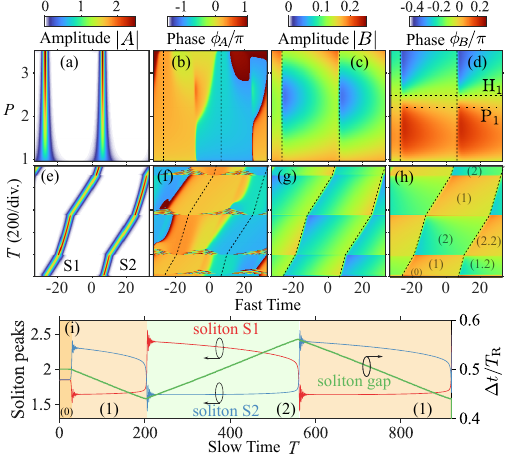}	
	\caption{
        (a-d) Field evolution of $|A|$, $\phi_A$, $|B|$, and $\phi_B$ for the PDSC on branch (iii) in Fig.~\ref{fig2_diagram}(a), as the driving $P$ is varied with fixed $\delta = 2$.  
        (e-h) Time evolution of the corresponding fields in the four subfigures, initialized from the solution marked as $\rm P_1$ at $P = 2.2$ in (d). 
		Temporal evolution of soliton peaks and their separation $\Delta t/T_R$ are shown in (i).
        Dashed curves in (b-d,e-h) indicate the positions of the two soliton peaks.
        In (h) and (i), the dynamical regimes associated with state switching are labeled by (0), (1), (2), (1.2), and (2.2).
	}
	\label{fig3_pump_phase}
\end{figure}

Notably, due to the phase-selective nature of PDCS, each soliton can be either in-phase or out-of-phase. Thus, in all cases shown in Fig.~\ref{fig2_diagram}(i–vii), solitons can adopt either phase state during excitation without affecting PDSC stability. Consequently, the branches in Fig.~\ref{fig2_diagram}(a) encompass all possible PDSC phase configurations, whose impact will be discussed later In this work.

We have examined the case of $\delta = 2$, but how do the formation and stability of PDSC change for different values of $\delta$?
Figure~\ref{fig2_diagram}(b) presents a phase diagram of PDSC as a function of $P$ and $\delta$. For clarity, we highlight only the region where stable PDSC states exist, with more details on fold and Hopf bifurcations provided in Appendix~\ref{sec_bifur_SC}, {\color{black} which indicates the transition between the PDSC states.} PDSC occupy the regions corresponding to different soliton numbers, with higher soliton counts requiring larger pump strengths. Some regions overlap, indicating bistability in PDSC.  
Notably, most soliton crystals are confined to regions where $\delta < 4$. 
This limitation arises because, for a fixed driving $P$, increasing $\delta$ amplifies PDSC peak intensities, leading to greater energy transfer from the pump to the signal. Consequently, the pump becomes more depleted, leaving insufficient energy to sustain PDSC at higher detuning. 
The maximum number of solitons is constrained by the cavity round-trip time and soliton duration, allowing up to six solitons in the current configuration. Therefore, increasing the normalized round-trip window $T_\mathrm{R}$, while keeping the remaining normalized parameters comparable, can accommodate a larger number of well-separated solitons.

\begin{figure*}[!t]
	\centering
	\includegraphics[scale=1]{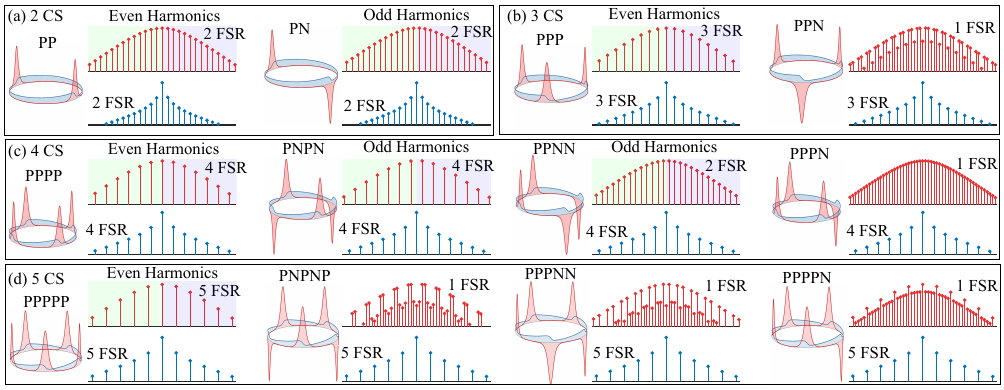}
	\caption{
		Simulations of PDSC with various phase configurations and their corresponding frequency combs. (a) Two CS with phase configurations PP (both positive) and PN (one positive, one negative). (b) Three CS with configurations PPP and PPN. (c) Four CS with configurations PPPP, PNPN, PPNN, and PPPN. (d) Five CS with configurations PPPPP, PNPNP, PPPNN, and PPPPN. 
	}
	\label{fig4_combs}
\end{figure*}

{\it Pump field.}
PDSC deplete the pump field consistently across all configurations. The evolution of the peak pump field $\mathrm{max}(|B|)$ is shown by the blue or gray curves in Fig.~\ref{fig2_diagram}(a). Unlike the unstable HS, where $\mathrm{max}(|B|)$ (gray curve) increases with $P$, stable PDSC deplete $B$, confining it within $0.15 < |B| < 0.2$.  
This indicates that although increasing $P$ raises the peak values of PDSC, the pump field reaches a threshold beyond which additional side peaks emerge on the leading side of each soliton due to excess OPO gain from the intensified pump. Consequently, this gain limitation regulates PDSC formation, with higher driving $P$ enabling the formation of PDSC with a greater number of solitons. 
{ 
Pump depletion drives soliton motion, analogous to gain depletion in harmonic mode-locking lasers (HML) \cite{Kutz1998,Horowitz2000,Gumenyuk2020,Liu2019,Ribenek2021}. Whereas disordered pulse phases in HML can produce incoherent soliton crystal states \cite{julien2025phase}, PDSCs remain phase-locked to the driving field, which enforces both equal spacing and precise phase alignment, yielding robust, highly coherent crystals.}

Different pump field phase states distinctly determine PDSC stability.
Figure~\ref{fig3_pump_phase}(a--d) plots the temporal amplitude and phase distributions of the signal and pump fields for the solutions on branch (iii) of Fig.~\ref{fig2_diagram}(a). The amplitude $|A|$ increases with driving $P$ [see Fig.~\ref{fig3_pump_phase}(a)], while the phases $\phi_A$ at CS center remain nearly unchanged, maintaining an anti-phase relationship [see Fig.~\ref{fig3_pump_phase}(b)]. At the soliton positions, the pump distribution $|B|$ exhibits a characteristic sharp decrease in its trailing edge [see Fig.~\ref{fig3_pump_phase}(c)], while the phase $\phi_B$ undergoes distinct shape transitions, separated by the Hopf bifurcation $\rm H_1$ [dashed line in Fig.~\ref{fig3_pump_phase}(d)]. 
Below this bifurcation line, PDSC are unstable. 
Notably, this line also marks a transition in pump phase behavior: for $P$ above this threshold, the pump phase undergoes a steep drop at the soliton’s peak position, whereas for lower $P$, it rises instead. This behavior emphasizes the critical role of the pump phase in stabilizing PDSC. 

Conventional cavity solitons can be trapped or guided by phase modulation of the background \cite{Sun2022OL,Sun2023chaos,Sun2023transition,englebertDynamicsDrivenDissipative2026}, as well as by amplitude and phase modulations of the driving field~\cite{Erkintalo2022,Talenti2023}. Likewise, the phase of pump $B$ plays a crucial role in stabilizing PDSC. In the stable region, the pump phase attracts solitons to positions with higher pump fields where they experience enhanced OPO gain.
This interplay between amplitude and phase ensures equalized soliton spacing, a critical condition for stable PDSC. Otherwise, solitons in PDSC may bifurcate into different states.

To illustrate this, we use the solution when $P=2.2$ at $\rm P_1$ (dashed line in Fig.~\ref{fig3_pump_phase}(d)) as the initial condition and simulate its time evolution. The temporal dynamics are shown in Fig.~\ref{fig3_pump_phase}(e-h). Initially, the solitons evolve symmetrically [Fig.~\ref{fig3_pump_phase}(e)] with an anti-phase relationship [Fig.~\ref{fig3_pump_phase}(f)], while the pump amplitude [Fig.~\ref{fig3_pump_phase}(g)] and phase [Fig.~\ref{fig3_pump_phase}(h)] remain unchanged in regime (1). The peak amplitudes of solitons S1 and S2, along with their normalized temporal separation $\Delta t/T_\mathrm{R}$, are plotted in Fig.~\ref{fig3_pump_phase}(i).  
At $T \approx 25$, symmetry breaking occurs in process (0), leading to process (1) [marked in Fig.~\ref{fig3_pump_phase}(h) and (i)]: S1 abruptly decreases in amplitude, while S2 undergoes amplification [see also 
Movie \href{https://youtu.be/cR1W5fUD8-k}{V}]. Simultaneously, the pump phase shifts, lowering the in-phase region (1.2) and increasing the relative phase in region (1). 
This causes S1 to decelerate (larger group delay) while S2 accelerates, thereby reducing the normalized soliton separation $\Delta t/T_\mathrm{R}$ [see the green curve in Fig.~\ref{fig3_pump_phase}(i)].
As the solitons move closer, S1 undergoes amplification while S2 experiences attenuation. By $T \approx 200$, they exchange intensities, transitioning to process (2): the pump phase flips, forcing the solitons to move apart [see Fig.~\ref{fig3_pump_phase}(h)]. 

{\it Soliton-pursuing state}. This periodic alternation persists throughout propagation, a unique phenomenon we term the {\it soliton-pursuing} state, where solitons periodically exchange intensities and group velocities.  
Notably, due to the cavity’s circular nature, the divergence (1) and convergence (2) processes are equivalent. 
Increasing the driving $P$ stabilizes these interactions by flipping phases [see Fig.~\ref{fig3_pump_phase}(h)], while decreasing $P$ extends the periodic cycles [see examples in 
Movie \href{https://youtu.be/QpL-oRu6uGM}{VI}, \href{https://youtu.be/69NMGB0B1EU}{VII} 
and the three-soliton pursuing state in 
Movie \href{https://youtu.be/atkTpMimGkU}{VIII}]. 
{\color{black}
From a bifurcation viewpoint, this regime results from a Hopf bifurcation of the soliton-crystal branches [e.g., H$_1$ in Fig.~\ref{fig2_diagram}(a)], yielding a time-periodic attractor where the relative soliton separations and peak powers oscillate on a time scale much longer than the cavity round-trip. Unlike standard breather solitons that oscillate at a fixed position, the PDSC group velocity depends on the driving conditions [see Figs.~\ref{fig2_diagram_d1} and \ref{fig_scan_pump} in Appendix], so individual solitons drift at different velocities while they oscillate.
}

{\it Soliton phases and corresponding combs.}  
OFC formed by conventional SC can be influenced by defects \cite{Hu2024,Cole2017}, such as missing solitons or quasi-periodicity within the crystal. 
In our case, PDSC are defect-free, as the soliton temporal separation is equalized by the pump fields.  
However, due to the phase-selective nature of OPOs, each soliton within PDSC branches (i–vii) can exist in two phase states. We now examine how these soliton phases affect the resulting OFC.

For a single soliton, the phase can be either positive (P) or negative (N), which does not affect the OFC. However, as the number of solitons increases, the OFC becomes strongly dependent on the relative soliton phases. Figure~\ref{fig4_combs} summarizes OFC formed by PDSC states (of Fig.~\ref{fig2_diagram}(i–vi)) for different soliton numbers and binary phases.  

For two solitons [Fig.~\ref{fig4_combs}(a)], the system converges to two degenerate phase configurations, PP and PN.
Both yield a doubled line spacing (2-FSR) with a $\mathrm{sech}^2$ envelope, but their spectral parity differs by symmetry. Specifically, for equally spaced solitons separated by $T_R/2$, the in-phase state satisfies $A(t+T_R/2)=A(t)$, which cancels all odd cavity modes and leaves only even harmonics, whereas the out-of-phase state satisfies $A(t+T_R/2)=-A(t)$, suppressing even modes and producing an odd-harmonic comb.  
In contrast, the pump comb remains even-harmonic for both PP and PN because the parametric coupling depends on $A^2$, which is invariant under $A\rightarrow -A$; thus the pump depletion retains a $T_R/2$ periodicity.
For three solitons [Fig.~\ref{fig4_combs}(b)], there are only two degenerate cases: PPP and PPN. 
In PPN, the components at $3n\Omega$ are reduced relative to the other two groups of comb lines.
A general derivation for an $N$-soliton crystal with arbitrary binary phase patterns is given in Appendix.

Interestingly, the soliton number $N$ fixes the \emph{temporal lattice} (equal spacing), but it does not uniquely fix the \emph{comb line spacing} of the signal spectrum.
The latter is determined by the discrete translation symmetry of the \emph{complex} field $A(t)$, which depends on the binary phase pattern across the $N$ lattice sites.
If all solitons share the same phase, $A(t)$ repeats after $T_R/N$ and the signal comb contains only every $N$th cavity mode (an $N$-FSR comb).
However, the phase pattern breaks the $N$-fold translational symmetry, which may yield \emph{subharmonic-like} spacings.
This is illustrated most clearly for $N=4$ in Fig.~\ref{fig4_combs}(c).
For four solitons, PPPP is periodic under a $T_R/4$ translation and therefore produces the mode family $k=4m$. The alternating PNPN pattern has the translation $A(t+T_R/4)=-A(t)$, and produces the shifted family $k=4m+2$. Likewise, PPNN is under a $T_R/2$ translation, suppressing all even cavity modes and leaving the odd-mode family $k=2m+1$. In contrast, PPPN has no nontrivial translation or anti-translation symmetry and therefore restores the full 1-FSR cavity-mode grid.

This symmetry-controlled ``frequency-division'' behavior is directly analogous to discrete time-translation symmetry breaking in discrete time crystals \cite{zhangObservationDiscreteTime2017,yaoDiscreteTimeCrystals2017}, where a system driven with period $T$ responds at an integer multiple $nT$.
Here, the pump comb enforces a $T_R/4$-periodic drive for the four-site lattice, while specific binary phase patterns yield signal-field responses with effective periods $T_R/2$ or $T_R$, corresponding to 2-FSR or 1-FSR line spacings---i.e., subharmonics of the 4-FSR pump.

More complex cases, such as the five-soliton scenario, are shown in Fig.~\ref{fig4_combs}(d). 
The generalized spectrum is related to the discrete Fourier transform of $\{(-1)^{p_n}\}$, where $p_n$ is the $n$-th soliton phase [see details in Appendix]. 
For even-$N$ PDSCs, odd-harmonic or parity-shifted combs require equal numbers of P and N solitons, whereas odd-$N$ PDSCs generally produce DFT-weighted 1-FSR combs rather than strictly odd-harmonic spectra.

{\it Conclusion.} 
We have theoretically characterized parametrically driven soliton crystals in doubly resonant cavities with quadratic and cubic nonlinearities. Pump–signal walk-off, together with pump depletion and pump phase, turns multi-soliton states into stable attractors: depletion becomes a long-range interaction that stabilizes equal spacing, while the driving strength primarily sets the soliton number. We also identify a new dynamical regime, the {\it soliton-pursuing} state, in which solitons periodically exchange intensities and group velocities, distinct from conventional breathing dynamics. 
Finally, we show that, once a binary phase pattern is realized, it maps deterministically onto the comb-mode weights and can generate odd-mode, shifted-harmonic, and subharmonic-like spectra. Controlled preparation of these phase patterns could therefore provide a route toward programmable spectral selection.
}

This work was supported by Marie Skłodowska-Curie Actions (101149506, 101150387, 101202061); ERC Consolidator Grant (101125625); F.R.S.-FNRS Chargé de recherches (40010332, 40031259); FWO and F.R.S.-FNRS under the Excellence of Science (EOS, 40007560).


\bibliography{references}

\appendix
\setcounter{figure}{0}
\renewcommand{\thefigure}{S\arabic{figure}}

\setcounter{table}{0}
\renewcommand{\thetable}{S\arabic{table}}

\setcounter{equation}{0}
\renewcommand{\theequation}{S\arabic{equation}}

\section{Model normalization.}  

The governing equations with physical units are:  
\begin{equation}
	\begin{aligned} 
		\frac{\partial \tilde{A}}{\partial \tilde{T}} 
		+&\frac{\beta_\mathrm{c}L}{T_R}\frac{\partial\tilde{A}}{\partial \tilde{t}}					
		= \left[ 
		-\frac{\alpha_1}{2}
		-i\tilde{\delta}
		-i\frac{\beta_{21}L}{2T_R}\frac{\partial^2}{\partial \tilde{t}^2}
		\right]\tilde{A}
		\\&+i\frac{L}{T_R}
		\left[
		(\gamma_1|\tilde{A}|^2+2\gamma_1|\tilde{B}|^2)\tilde{A}+\tilde{g}\tilde{B}\tilde{A}^*
		\right],	
	\end{aligned}
	\label{eqA_unit}
\end{equation}
\begin{equation}
	\begin{aligned} 
		&\frac{\partial \tilde{B}}{\partial \tilde{T}} 
		+\frac{\beta_\mathrm{c}L}{T_R}\frac{\partial\tilde{B}}{\partial \tilde{t}}						
		= \left[ -\frac{\alpha_2}{2}
		-i\left(2\tilde{\delta}+\frac{\tilde{\Delta \beta_0} L}{T_R}\right)
		-i\frac{\beta_{22}L}{2T_R}\frac{\partial^2}{\partial \tilde{t}^2}\right]\tilde{B}
		\\-&\frac{\tilde{\Delta \beta_1}L}{T_R}\frac{\partial\tilde{B}}{\partial \tilde{t}}+i\frac{L}{T_R}\left[
		(\gamma_2|\tilde{B}|^2+2\gamma_2|\tilde{A}|^2)\tilde{B}+\tilde{g}\tilde{A}^2
		\right]
		+\sqrt{\frac{\kappa}{T_R}}B_{\rm in},		
	\end{aligned}
	\label{eqB_unit}
\end{equation}
where $\tilde{T}$ and $\tilde{t}$ are slow and fast times, respectively; $T_R$ is the cavity roundtrip time; $\tilde{\delta}$ is the detuning; $\tilde{\Delta \beta_0}$ is the phase mismatch; $\beta_{21}$ and $\beta_{22}$ are the group velocity dispersions for signal and pump; $\alpha_1$ and $\alpha_2$ represent signal and pump losses; $\gamma_1$ and $\gamma_2$ are the nonlinear coefficients for signal and pump; $\tilde{g}$ is the quadratic coefficient; $L$ is the cavity length; $\kappa$ is the coupling rate; and $B_{\rm in}$ is the driving field. 
To improve visualization and analysis, we add a group delay term $\frac{\beta_\mathrm{c}L}{T_R} \frac{\partial}{\partial \tilde{t}}$ in the moving reference frame for both fields, which does not affect the underlying results. 

For clarity, notations with a tilde represent physical quantities, while the same notations without a tilde correspond to their dimensionless, normalized counterparts.
The normalized equations are:
\begin{equation}
	\begin{aligned} 
		\frac{\partial A}{\partial T} 
		+d_c\frac{\partial A}{\partial t}					
		&= \left[ 
		-1
		-i\delta
		+i\frac{\partial^2}{\partial t^2}				
		\right]A
		\\&+
		i\left[
		(\left|A\right|^2+2|B|^2)A
		+gBA^*\right],
	\end{aligned}
	\label{eq_A4}
\end{equation}				
\begin{equation}
	\begin{aligned} 
		\frac{\partial B}{\partial T} 
		+d_c\frac{\partial B}{\partial t}			
		&= \left[ -r_\mathrm{l}
		-i\left(2\delta+\delta_0\right)
		+ir_\mathrm{g}\frac{\partial^2}{\partial t^2}
		-d\frac{\partial}{\partial t}
		\right]B
		\\&+i\left[
		r_\mathrm{n}(\left|B\right|^2
		+2\left|A\right|^2)B
		+gA^2
		\right]
		+P,
	\end{aligned}	
	\label{eq_B4}
\end{equation}
where the normalized parameters are summarized in Tab.~\ref{tab_norm}. Using physical parameters $L = 0.4\,\rm mm$, $T_R = 2.8\,\rm ps$, $\alpha_1 = \alpha_2/4 = \kappa/2 = 2\pi \times 0.65\,\rm GHz$, $\tilde{g} = 40\,\rm W^{-1/2}m^{-1}$, $\gamma_1 = 0.75\,\rm (Wm)^{-1}$, $\gamma_2 = 1.5\,\rm (Wm)^{-1}$, $\beta_{21} = -50\,\rm fs^2/mm$, $\beta_{22} = 50\,\rm fs^2/mm$, $\tilde{\Delta\beta_1} = 0.4\,\rm ps/mm$, the normalized values are derived as shown in the Letter. The normalized signal-dispersion term becomes \(+i\partial_t^2A\) for the anomalous-dispersion case \(\beta_{21}<0\), which is the regime considered here. The pump-dispersion coefficient is then \(r_{\rm g}=\beta_{22}/\beta_{21}\), so that the normalized term \(ir_{\rm g}\partial_t^2B\) recovers the dimensional coefficient \(-i\beta_{22}L/(2T_R)\).
In addition, $\tilde{\Delta\beta_1}=\beta_{1,2\omega}-\beta_{1,\omega}$, so that $\tilde{\Delta\beta_1}>0$ means that the second-harmonic pump field has a larger group delay than the fundamental signal field.

\begin{table}[h!]
	\centering
	\caption{Normalization for fields and physical parameters.}
	\label{tab:normalization}
	\renewcommand{\arraystretch}{1.5}
	\begin{tabular}{|c|c|}
		\hline
        $\beta_1 = \frac{T_R}{L}$
        & $r_\mathrm{g} = \frac{\beta_{22}}{\beta_{21}}$ \\ \hline
        $A_\mathrm{s} = \sqrt{\frac{\alpha_1T_R}{2L\gamma_1}}= \sqrt{\frac{\alpha_1\beta_1}{2\gamma_1}}$   
        & $r_\mathrm{l} = \frac{\alpha_2}{\alpha_1}$ \\ \hline        
		$t_\mathrm{s} = \sqrt{\frac{|\beta_{21}|L}{\alpha_1T_R}}= \sqrt{\frac{|\beta_{21}|}{\alpha_1\beta_1}}$ 		
        & $r_\mathrm{n} = \frac{\gamma_2}{\gamma_1}$ \\ \hline
		$T_\mathrm{s} = \frac{2}{\alpha_1}$ 
        & $\delta = \tilde{\delta}\frac{2}{\alpha_1} = \tilde{\delta} T_\mathrm{s}$ \\ \hline
		$A = \frac{\tilde{A}}{A_\mathrm{s}}$ 
        & $\delta_0 = \tilde{\Delta \beta_0}\frac{2L}{\alpha_1T_R} = \tilde{\Delta \beta_0}\frac{1}{\gamma_1A^2_\mathrm{s}}$ \\ \hline
		$B = \frac{\tilde{B}}{A_\mathrm{s}}$ 
        & $d = \tilde{\Delta \beta_1}\sqrt{\frac{4L}{\alpha_1|\beta_{21}|T_R}} 
		= \tilde{\Delta \beta_1} \frac{2t_\mathrm{s}}{|\beta_{21}|}$ \\ \hline
		$T = \frac{\tilde{T}}{T_\mathrm{s}}$ 
        & $d_c = \beta_\mathrm{c} \frac{2t_\mathrm{s}}{|\beta_{21}|}$ \\ \hline
		$t = \frac{\tilde{t}}{t_\mathrm{s}}$ 
        & $g = \tilde{g}\sqrt{\frac{2L}{\alpha_1\gamma_1T_R}} 
		= \tilde{g}\frac{1}{\gamma_1A_\mathrm{s}}$ \\ \hline
		& $P = B_{\rm in}\sqrt{\frac{8L\kappa\gamma_1}{\alpha_1^3T_R^2}}
= B_{\rm in}\sqrt{\frac{\kappa}{T_R}}\frac{T_\mathrm{s}}{A_\mathrm{s}}$ \\ \hline
	\end{tabular}
	\label{tab_norm}
\end{table}

\section{Comoving-frame drift coefficient}	

Here, in Fig.~\ref{fig2_diagram_d1}, we show the evolution of the moving frame delay corresponding to each branch in Fig.~2(a) in the Letter. For each branch, as the driving strength $P$ increases, the moving frame delay $d_\mathrm{c}$ decreases, indicating an increase in the soliton crystal's group velocity. 

\begin{figure}[h!]
	\centering
	\includegraphics[scale=1]{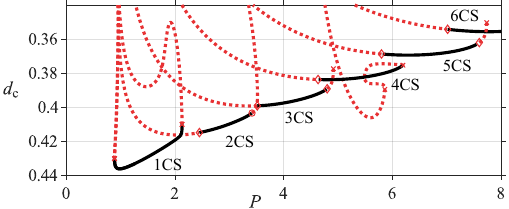}	
	\caption{
		Bifurcation diagram showing the moving frame delay $d_\mathrm{c}$ for the soliton crystal, with black and red curves representing stable and unstable solutions, respectively, as a function of driving strength $P$. Positive values of $d_\mathrm{c}$ indicate that, in the original reference frame, the soliton crystal drifts toward smaller fast time, corresponding to a reduced group delay and thus a higher group velocity. 
	}
	\label{fig2_diagram_d1}
\end{figure}

This behavior is consistent with the large pump-signal walk-off, characterized by $d=669$. With our convention, $d>0$ means that the second-harmonic pump field has a larger group delay than the fundamental signal field; equivalently, the fundamental field propagates faster than the pump field. As the driving strength increases, the signal power and soliton number grow, while the background remains nearly unchanged, collectively increasing the soliton crystal's group velocity.

\section{Soliton excitation by detuning scanning.}\label{sec_excitation_scandetuing}

In comparison with Fig.~1 in the main text, Figure~\ref{fig1_sup_concept_excitation} shows PDSC formation with detuning scans for $P = 2.5$ [Figs.~\ref{fig1_sup_concept_excitation}(a–c)] and $P = 6.5$ [Figs.~\ref{fig1_sup_concept_excitation}(d–f)], starting from white noise ($\max|A| = 0.01$). The detuning $\delta$ is linearly swept from $-2$ to $2$ over $T = 100$ and then held constant.  
For $P = 2.5$, the temporal evolution of $|A|$ [Fig.~\ref{fig1_sup_concept_excitation}(a)] and $|B|$ [Fig.~\ref{fig1_sup_concept_excitation}(b)]
is chaotic for $0 < T < 100$, then stabilizes into two solitons. At $T = 1000$, the steady state [Fig.~\ref{fig1_sup_concept_excitation}(c)] features a sawtooth pump and $\mathrm{sech}$-shaped solitons aligned with pump edges.  
For $P = 6.5$, five PDSCs form [Figs.~\ref{fig1_sup_concept_excitation}(d–f)], initially irregular but converging to evenly spaced, uniform-intensity patterns. 
In both cases, detuning scans lead to PDSCs with dynamics qualitatively similar to those obtained at fixed detuning (Fig.~1 in the main text).

\begin{figure}[h!]
	\centering
	\includegraphics[scale=1]{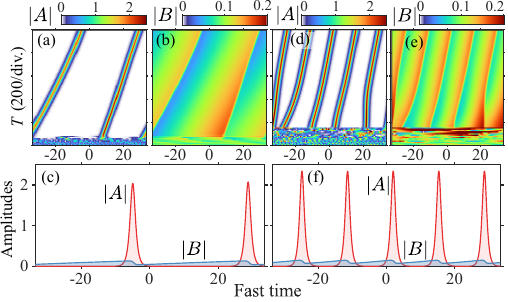}
	\caption{
Evolution of $|A(t,T)|$ and $|B(t,T)|$ over time, along with the final profile $|A(t,T=1000)|$, for driving strengths $P = 2.5$ in (a-c) and $P = 6.5$ in (d-f). The detuning parameter $\delta$ is linearly scanned from $-2$ to $2$ over the interval $T = 0$ to $T = 100$. (c,f) Final temporal profiles at $T = 1000$.
	}
	\label{fig1_sup_concept_excitation}%
\end{figure}

{An additional group delay $d_\mathrm{c}$ was applied here in the reference frame for both fields to keep solitons centered for clearer visualization, without affecting results: $d_\mathrm{c} = 0.415$ in Fig.~\ref{fig1_sup_concept_excitation}(a-c) and $d_\mathrm{c} = 0.352$ in Fig.~\ref{fig1_sup_concept_excitation}(d-f). The two- and five-soliton buildup processes are shown in Movies~\href{https://youtu.be/RJua9gek3ig}{I} and Movie~\href{https://youtu.be/qzHT1moY9Ys}{II}, respectively.}

\section{Evolution of soliton crystals via scanning driving strength}

To further confirm the connection between the dynamically observed soliton crystals and the steady-state branches shown in Fig.~2(a) of the Letter, we simulate the evolution of a six-soliton crystal while gradually decreasing the driving strength. The simulation starts from a stable six-soliton crystal at $P=7.5$ and $\delta=2$, with all other parameters identical to those used in Fig.~2(a) of the Letter. As $P$ is reduced, the soliton crystal undergoes a sequence of transitions in which individual solitons disappear one by one. Consequently, the soliton number decreases stepwise, passing through the multi-soliton states identified in the bifurcation diagram. This dynamical scan therefore provides an independent time-domain confirmation of the existence of the soliton-crystal branches.

\begin{figure}[!h]
	\centering
    \includegraphics[scale=1]{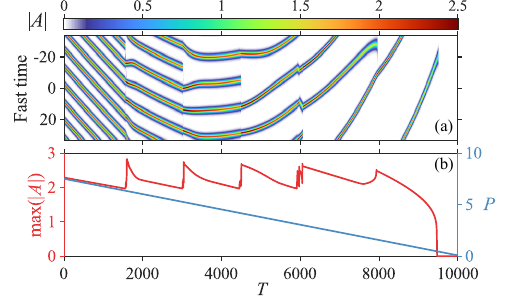}		
	\caption{
Temporal evolution of $|A|$ while the driving strength $P$ is gradually reduced from 7.5 to 0 at fixed detuning $\delta=2$ [see Movie~\href{https://youtu.be/-vWOQzNRxHg}{IV}]. 
(a) Evolution of the intracavity field amplitude $|A|$. 
(b) Peak amplitude $\max(|A|)$ extracted from (a), together with the applied driving strength $P$. 
The soliton number decreases stepwise as the driving is reduced, illustrating the sequential transitions between soliton-crystal states. Other parameters are the same as in Fig.~2(a) of the Letter.
	}
	\label{fig_scan_pump}
\end{figure}

\section{Bifurcation and phase diagram for homogeneous solutions}\label{sec_homo}

Here, we present the bifurcation and phase diagrams of the homogeneous solutions (HS) of the normalized equations in the Letter. 

Figure~\ref{homo_bif_delta_2p0_P__3p0_01}(a) shows the bifurcation diagram of $|A|$ (red) and $|B|$ (blue) as functions of $P$ with $\delta = 2$. In the low-$P$ regime, only one type of solution exists: the pump field $|B|$ increases with $P$, while the signal field $|A| = 0$. This is classified as HS type I, which remains stable over a range of $P$ until it reaches the bifurcation point $\rm B_2$ at approximately $P = 1$. At $\rm B_2$, a new branch, HS type II, emerges, where the signal field $|A|$ begins to increase while the pump field $|B|$ decreases. Although this branch crosses the fold bifurcation $\rm F_2$, all parts of the HS type II branch are unstable.

\begin{figure*}[!th]
	\centering
	\includegraphics[scale=1]{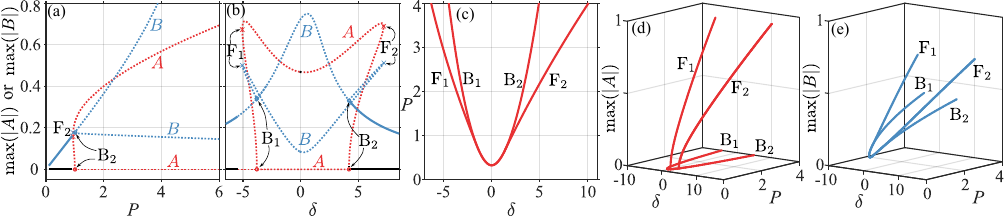}		
	\caption{ 
		Bifurcation of $|A|$ (red) and $|B|$ (blue) as a function of $P$ when $\delta=2$ in (a) and as a function of $\delta$ when $P=3$ in (b);	 
		(c,d,e) Phase diagram of  bifurcations ($\rm F_1,\,F_2,\,B_1,\,B_2$) of $|A|$ (red) or $|B|$ (blue) as a function of $\delta$ and $P$. Other parameters are the same as those in the Letter.		
	}
	\label{homo_bif_delta_2p0_P__3p0_01}
\end{figure*}

We further trace these solution branches by fixing $P = 3$ and varying $\delta$, as illustrated in Fig.~\ref{homo_bif_delta_2p0_P__3p0_01}(b). The HS type I solution is stable for both large and small detuning $\delta$, with $|A| = 0$. In the negative $\delta$ regime, $|B|$ increases with $\delta$, while in the positive $\delta$ regime, $|B|$ begins to decrease as $\delta$ increases. For the HS type II branch, where $|A| \neq 0$, solutions appear at either $\rm B_1$ or $\rm B_2$ and exhibit fold bifurcations at $\rm F_1$ or $\rm F_2$.

These bifurcations are summarized in the phase diagram shown in Fig.~\ref{homo_bif_delta_2p0_P__3p0_01}(c,d,e). The bifurcation points $\rm B_1$ and $\rm B_2$ occur at $(P, \delta) \approx (0.33, 0)$. For $P \lesssim 0.33$, only one stable homogeneous solution exists. The fold bifurcations $\rm F_1$ and $\rm F_2$ appear at $(P, \delta) \approx (0.65, 1.36)$ and $(P, \delta) \approx (0.65, -1.28)$.

\section{Bifurcation and phase diagrams for PDSC.}\label{sec_bifur_SC}

\begin{figure*}[!th]
	\centering
	\includegraphics[scale=1]{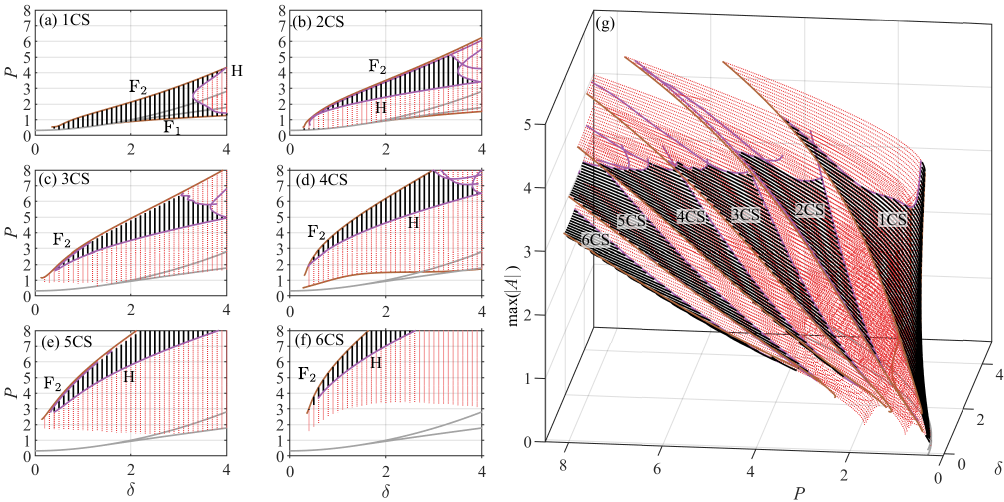}			
	\caption{Phase diagrams detailing fold $\rm F_2$ (brown curves) and Hopf $\rm H$ (purple curves) bifurcations for soliton crystals with 1 to 6 solitons are shown in (a–f). The gray curves represent the fold bifurcations of homogeneous solutions. The evolution of the maximum amplitude $|A|$ for soliton crystals is depicted in (g).
	}
	\label{fig_SC_phase_dia}
\end{figure*}

In Fig.~\ref{fig_SC_phase_dia}, we provide detailed phase diagrams of fold (brown curves) and Hopf (purple curves) bifurcations for soliton crystals with 1 to 6 solitons in (a–f). The gray curves indicate the fold bifurcations of homogeneous solutions. The peak values of soliton crystals on the phase diagram are shown in (g). Black regions represent the stable SC regimes, corresponding to Fig.~2(b) in the Letter.

\section{Soliton states affected by walk-off $d$.}\label{sec_walkoff}

To explore how localized soliton states are influenced by the walk-off delay between the pump and signal fields, Fig.~\ref{fig_walk_off}(a) presents the phase diagram of fold bifurcations $\rm F_1$ and $\rm F_2$ as functions of $d$ and $\delta$ for $P=1$ and $P=2$. The bifurcation points $\rm F_1$ and $\rm F_2$ are the same as shown in Fig.~\ref{fig_SC_phase_dia}(a), delineating the regions where soliton states can exist.

\begin{figure}[!h]
	\centering
	\includegraphics[scale=1]{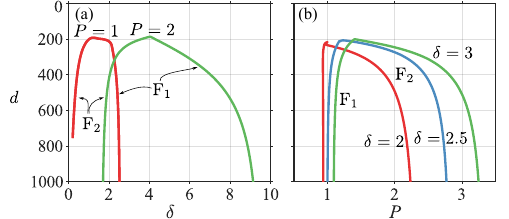}	
	\caption{Phase diagrams of single-soliton fold bifurcations $\rm F_1$ and $\rm F_2$: (a) Dependence on $d$ and $\delta$ for $P=1$ and $P=2$. (b) Dependence on $d$ and $P$ for $\delta=2.0$, $\delta=2.5$, and $\delta=3.0$.}
    \label{fig_walk_off}
\end{figure}

A similar phase diagram can be observed in Fig.~\ref{fig_walk_off}(b), where driving strength $P$ is varied for different detuning values $\delta$. The diagrams indicate that localized soliton states exist when $d \gtrsim 200$. This highlights the wide parameter space for walk-off delay that permits the existence of localized states, as thoroughly analyzed in the main text.

\section{The role of quadratic nonlinear coefficient $g$}\label{sec_chi2strength}
\begin{figure}[!h]
	\centering
	\includegraphics[scale=1]{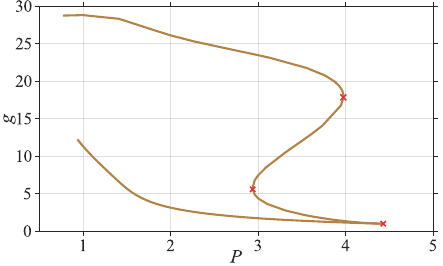}	
	\caption{Phase diagrams illustrating fold bifurcations $\rm F_1$ and $\rm F_2$ as functions of $P$ and $g$ for a two-soliton crystal at $\delta = 2$. All other parameters are identical to those in Fig.~\ref{fig_SC_phase_dia}(b).}
    \label{fig_SC_P_vs_g}
\end{figure}	

In Fig.~\ref{fig_SC_P_vs_g}, we conduct a general investigation into how the quadratic nonlinear coefficient $g$ impacts the nonlinear dynamical range of PDSC. In the main text, the normalized quadratic coefficient is $g = 12.2$. To further explore this, we analyze phase diagrams of fold bifurcations $\rm F_1$ and $\rm F_2$ as functions of $P$ and $g$ for a two-soliton crystal at $\delta = 2$. The results reveal that PDSC exists only within a limited range of $P$ and $g$. If $g$ is too large, solitons in the PDSC transition to other states.

\section{Frequency-comb selection rules for parametrically driven soliton crystals}

Here we summarize how the binary phases of parametrically driven soliton crystals determine the corresponding signal comb.
We consider an $N$-soliton crystal with equally spaced solitons within one cavity round trip $T_R$. The soliton positions are
\begin{equation}
t_n=
\left(
-\frac{N-1}{2N}+\frac{n}{N}
\right)T_R,
\qquad n=0,1,\ldots,N-1,
\end{equation}
and the cavity free spectral range is $\Omega=2\pi/T_R$.
Because the degenerate OPO process fixes each soliton phase to either $0$ or $\pi$, we write the relative phase of the $n$-th soliton as
\begin{equation}
s_n=(-1)^{p_n},\qquad
p_n=
\begin{cases}
0, & \mathrm{P},\\
1, & \mathrm{N}.
\end{cases}
\end{equation}
Up to an irrelevant global phase, the signal field can then be written as
\begin{equation}
E(t)=
\left[
\sum_{n=0}^{N-1}
s_n a(t-t_n)
\right]
*
\sum_{q=-\infty}^{\infty}\delta(t-qT_R),
\end{equation}
where $a(t)$ denotes the envelope of a single soliton. Its Fourier transform is
\begin{equation}
E(\omega)
=
\Omega a(\omega)
\sum_{k=-\infty}^{\infty}
\mathcal{C}_k
\delta(\omega-k\Omega),
\end{equation}
where the complex line-selection factor is
\begin{equation}
\mathcal{C}_k
=
e^{i\pi k (N-1)/N}
\sum_{n=0}^{N-1}
s_n e^{-i2\pi kn/N}.
\end{equation}
The prefactor $e^{i\pi k (N-1)/N}$ is an origin-dependent linear
spectral phase and does not affect the comb intensity. Therefore,
\begin{equation}
|E_k|^2
\propto
|a(k\Omega)|^2
\left|
\sum_{n=0}^{N-1}
s_n e^{-i2\pi kn/N}
\right|^2 .
\label{eq:DFT_comb_selection}
\end{equation}
Equation~\eqref{eq:DFT_comb_selection} shows that the signal spectrum is the single-soliton spectral envelope multiplied by the discrete Fourier transform of the binary phase sequence ${s_n}$. Thus, the soliton number fixes the temporal lattice, whereas the binary phase pattern determines which cavity modes survive.

Several useful cases follow directly from Eq.~\eqref{eq:DFT_comb_selection}. Phase patterns related by a global sign change or a cyclic shift have the same comb intensities.

\begin{table*}[!ht]
\centering
\caption{
Representative comb-selection rules for one- to four-soliton crystals. Here $k$ labels the cavity mode $\omega=k\Omega$, and only the magnitude of the DFT factor is shown because the remaining phase factors do not affect the comb intensity.
}
\begin{tabular}{c c c c}
\toprule
Soliton number & Phase pattern & Surviving cavity modes & DFT magnitude \\
\midrule
$N=1$ & P & all $k$ & $1$ \\
\midrule
$N=2$ & PP & $k=2m$ & $2$ \\
& PN & $k=2m+1$ & $2$ \\
\midrule
$N=3$ & PPP & $k=3m$ & $3$ \\
& PPN, or equivalent cyclic shifts & all $k$ & $|\mathcal{C}_{3m}|=1$, $|\mathcal{C}_{3m+1}|=|\mathcal{C}_{3m+2}|=2$ \\
\midrule
$N=4$ & PPPP & $k=4m$ & $4$ \\
& PNPN & $k=4m+2$ & $4$ \\
& PPNN & $k=2m+1$ & $2\sqrt{2}$ \\
& PPPN & all $k$ & $2$ \\
\bottomrule
\end{tabular}
\label{tab:comb_selection_rules}
\end{table*}

For two solitons, the PP and PN configurations both generate a 2-FSR comb, but with opposite parity: PP suppresses all odd cavity modes, while PN suppresses all even cavity modes. This explains the even- and odd-harmonic combs shown in Fig.~\ref{fig4_combs}(a).

For three solitons, the fully in-phase state PPP has a $T_R/3$ periodicity and therefore produces a 3-FSR comb. In contrast, a phase-defected state such as PPN breaks this threefold translation symmetry. As a result, all cavity modes are allowed, but their amplitudes are weighted by the DFT of the binary phase sequence; the modes $k=3m$ are weaker than the two other mode families.

For four solitons, the phase pattern gives a particularly clear example of symmetry-controlled frequency division. The PPPP state preserves the $T_R/4$ periodicity and produces a 4-FSR comb. The alternating PNPN state also gives a 4-FSR comb, but shifted to the mode family $k=4m+2$. The PPNN state has an effective repeat unit of $T_R/2$ and produces a 2-FSR odd-harmonic comb. Finally, PPPN breaks the discrete translation symmetry of the four-site lattice and restores a 1-FSR comb, while retaining the spectral envelope of the individual solitons.

This DFT picture also clarifies why even- and odd-$N$ crystals behave differently. For even $N$, phase patterns with equal numbers of P and N solitons have zero mean, $\sum_n s_n=0$, which suppresses the corresponding zero-order DFT component and enables parity-shifted or odd-harmonic comb structures. For odd $N$, exact cancellation between P and N solitons is impossible; phase-defected odd-$N$ crystals therefore generally produce DFT-weighted 1-FSR spectra rather than strictly odd-harmonic combs.

\end{document}